\documentstyle[11pt,newpasp,twoside, epsfig, epsf]{article}
\begin{document}
\title{The quasar 3C\,345 at the highest resolution with mm- and space-VLBI}
\author{J. Klare, J.A. Zensus, T.P. Krichbaum, A.P. Lobanov \& E. Ros}
\affil{Max-Planck-Institut f\"ur Radioastronomie, Auf dem H\"ugel 69, 53121 Bonn, Germany} 

\begin{abstract}
The quasar 3C\,345 is one of the best examples of extragalactic jet with superluminal components moving on helical paths. To get better constraints on the jet physics, we have imaged the source at the highest possible resolution, using mm-VLBI and space-VLBI observations. We observed 3C\,345 at 5 different epochs so far (1 epoch with 3-mm ground VLBI, and 4 epochs with space VLBI at 6\,cm and 18\,cm). These measurements expand significantly our ongoing monitoring of 3C\,345. In this contribution, we discuss the properties of the jet and its moving components in the immediate vicinity of the core ($<$ 3\,mas). 
\end{abstract}
\vspace{-0.8cm}
\section{Introduction}
The 16th magnitude quasar 3C\,345 (z=0.595) can be regarded as the archetypical source for studies of superluminal motion. Several partially resolved enhanced emission regions (components) are observed in the jet. They present strongly bent trajectories after having been ejected from the core at apparent speeds of 2$c$--20$c$ (e.g. Zensus, Cohen, \& Unwin 1995). The curvature of the trajectories can result from a periodic process driven by Kelvin-Helmholtz instabilities (Hardee 1987; Steffen et al. 1995; Qian et al. 1996) or a binary black hole system (Roos, Kaastra, \& Hummel 1993; Lobanov 1996). The ejection angles of the components vary, and the component trajectories differ significantly. The curvature of the component trajectories increases towards the core, while the trajectories appear to straighten at larger core separations. This makes the innermost region a better tool for studying the kinematics of the jet components. Thus, it is important to investigate their behaviour at a very early stage of their evolution. This makes 3C\,345 a prime candidate for high-resolution imaging with mm- and space-VLBI.
\vspace{-0.4cm}
\section{Observations and imaging}
We discuss here 4 VSOP observations at 6\,cm (1998.22, 1998.57, 1999.50, 1999.69) and one image from our ongoing 3-mm CMVA monitoring (1997.29) of 3C\,345. The VSOP data were processed at the NRAO\footnote{The National Radio Astronomy Observatory (NRAO) is operated by Associated Universities, Inc., under cooperative agreement with the National Science Foundation.} VLBA correlator in Socorro, New Mexico, USA, and the 3\,mm data were correlated at the MPIfR correlator in Bonn, Germany. Calibration and fringe-fitting were done in AIPS, and imaging was performed using DIFMAP. Model fitting by elliptical Gaussian components was applied to determine the flux densities and positions of the jet components.

\begin{figure}[h]
\begin{minipage}[b]{6.6cm}{
\psfig{figure=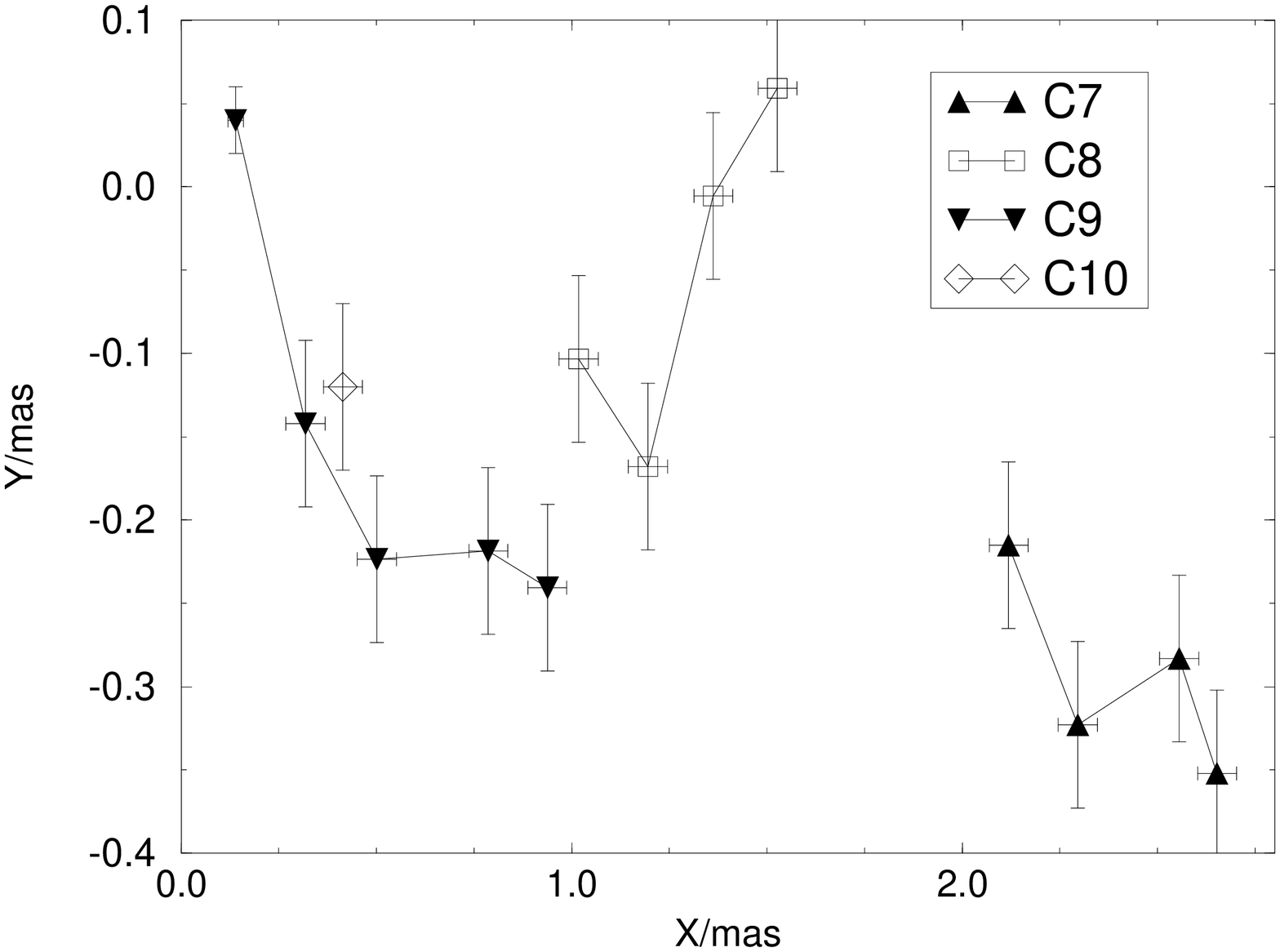,width=5cm,angle=-90} 
}
\end{minipage}
\begin{minipage}[b]{6.5cm}{
\psfig{figure=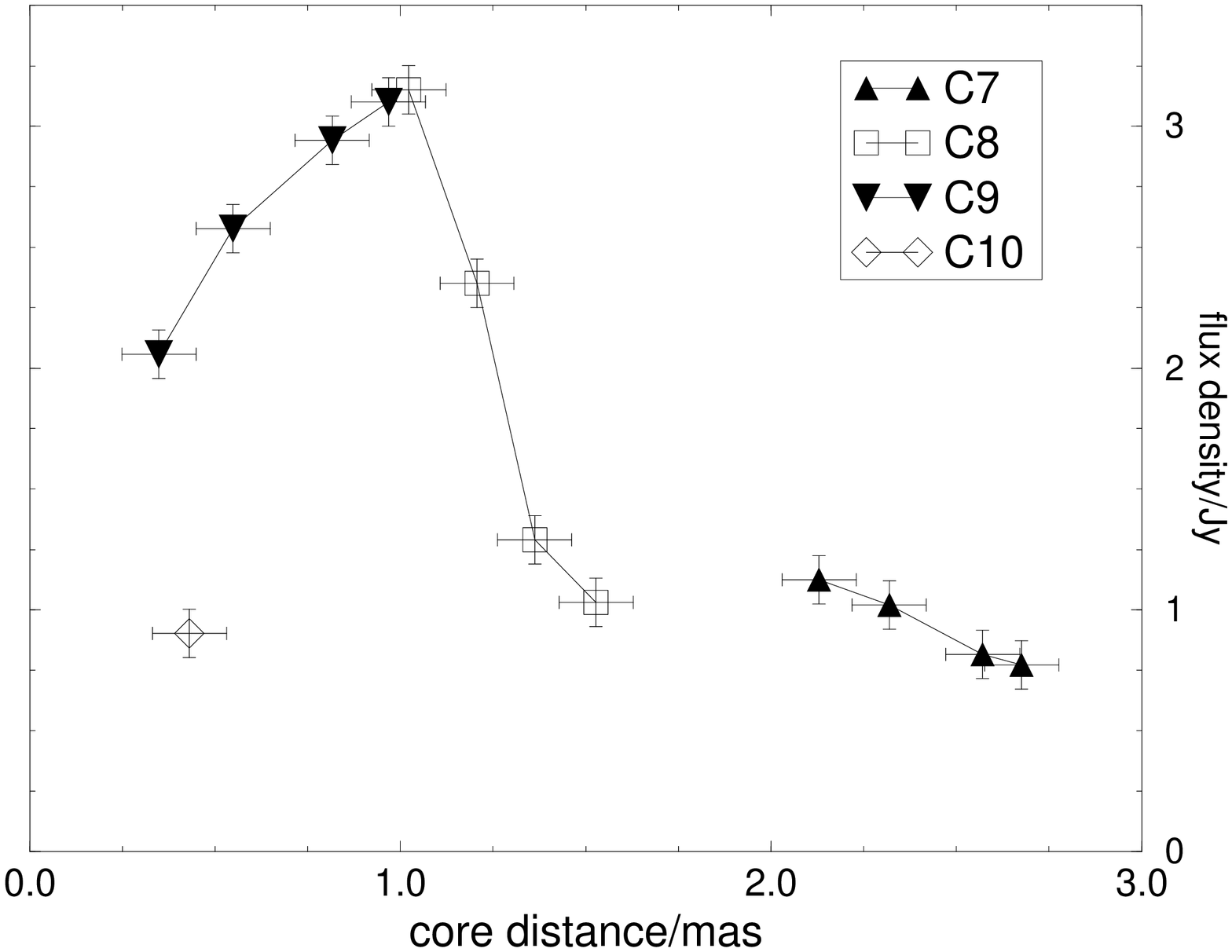,width=5cm,angle=-90}
}
\end{minipage}
\caption{Left: Trajectories of the jet components in respect of the core. Right: Flux evolution of the jet components.}
\end{figure}

\vspace{-0.8cm}
\section{The innermost jet components}
In the VSOP data at 6\,cm, the ground-space baselines are about 3 times longer than the longest ground baselines, leading to a threefold improvement of the angular resolution. At $\lambda$\,=\,6\,cm, the interferometric beam of the VSOP data is of {$\sim$ 0.35\,mas} in east-west direction, while the ground array yields only {$\sim$1\,mas} resolution. With this high image resolution, we are able to identify the synchrotron self-absorbed core and four inner jet components (C10, C9, C8 and C7) to the west of the core (Figure 2, top). In the ground array image of 3C\,345, the core and these components appear blended. To present and compare the four VSOP epochs, we have restored the images with a circular beam of 0.4\,mas, and modelled the brightness distribution by Gaussian components (Figure 2, top right). Significant changes of the component positions, flux densities, and sizes are evident in the images in Figure 2. VLBI observations at 3\,mm  provide an additional tool to trace the components at a resolution of 70$\mu$as in the jet direction (Figure 2, bottom), and determine the trajectories at the smallest distances from the core. Due to lower opacity of the emission, one can detect the components earlier in their evolution at 3\,mm than at longer wavelengths. The trajectories and the flux evolution of the components are shown in Figure 1.
\vspace{-0.1cm}
\subsection{Flux evolution}
The flux density of the core shows almost a threefold increase, during the 1.5-year period elapsed between the first and the last epoch. In 1999.69, when the core flux density reached 3\,Jy, we detected a new faint component C10 with 0.9\,Jy at a core separation of 0.43\,mas (Figure 1, right). C9, which was detected in 1998.22 at 0.35\,mas core separation, was more than 2 times brighter than C10. The flux density of C9 increased up to 3.1\,Jy at a core distance of 0.97\,mas. 1998.22 data show C8 at 1.0\,mas with a flux density of 3.2\,Jy. Later, its flux density has decreased down to 1.0\,Jy at 1.5\,mas core separation in 1999.69. For the jet-models, it will be very important to investigate with further observations if C9 would also decrease in flux density at core separations larger than 1\,mas. The flux density of C7 decreases slightly from 1.1\,Jy (1998.22) to 0.77\,Jy (1999.69) between 2.1\,mas and 2.7\,mas. C8 has already reached the 1.0\,Jy level at 1.5\,mas in 1999.69, and future observations should show if it will disappear quickly like C6 (Lobanov 1996) or only decrease slightly in flux density like C7 did.
\vspace{-0.1cm}
\subsection{Proper motions}
In the 3\,mm map, the position of C9 lies slightly to the north-west at a core distance of 0.14\,mas in 1997.29. The VSOP data show a trajectory of C9 to the south-west during the 1.5 year period. This means, within 0.14\,mas core distance the trajectory of C9 turned from north to south. C8 moved steep to the north-west from the second epoch on. The turn of the C8 trajectory from south to north near 1.2\,mas may cause the drop of flux density by changes of the Doppler boosting. C8 has an averaged angular speed of only 0.18\,mas/yr, which corresponds to an apparent speed of 3.6\,$h^{-1}c$ ($q_0=0.5$, $H_0=100h$\,km\,s$^{-1}$Mpc$^{-1}$). The components C9 and C7 show faster proper motions, of 0.3\,mas/yr (5.8\,$h^{-1}c$) and 0.28\,mas/yr (5.5\,$h^{-1}c$) in average. Also, the proper motion of C7 during 1996 was faster with 0.35\,mas/yr (6.8\,$h^{-1}c$) (Ros, Zensus, \& Lobanov 2000), at the same core separation as for C8. The observed relative weakness of C8 may be related to its lower apparent speed.
\vspace{-0.1cm}
\subsection{Conclusions and outlook}
Doppler boosting migth increase the flux density of C9 while it moves to the south-west and decrease the flux density of C8 while it moves to the north-west. This can be caused by helical trajectories of the jet components. The faster component C9 may merge with the slower component C8 in future observations. Moreover, the behavior of C8 and C9 is similar to C7 and C6 (Lobanov 1996), which may suggest that two ejected components are likely to be coupled to each other. With the follow up observations at 3\,mm, we will able to trace C9 and C10 and determine the strong curvatures of the trajectories close to the central engine. Our continuing effort in high resolution imaging, will help to get better constrains for the jet models of 3C\,345 (Klare et al. 2001).

\acknowledgements
We gratefully acknowledge the VSOP Project, which is led by the Japanese Institute of Space and Astronautical Science in cooperation with many organizations and radio telescopes around the world.

\begin{figure}
\psfig{figure=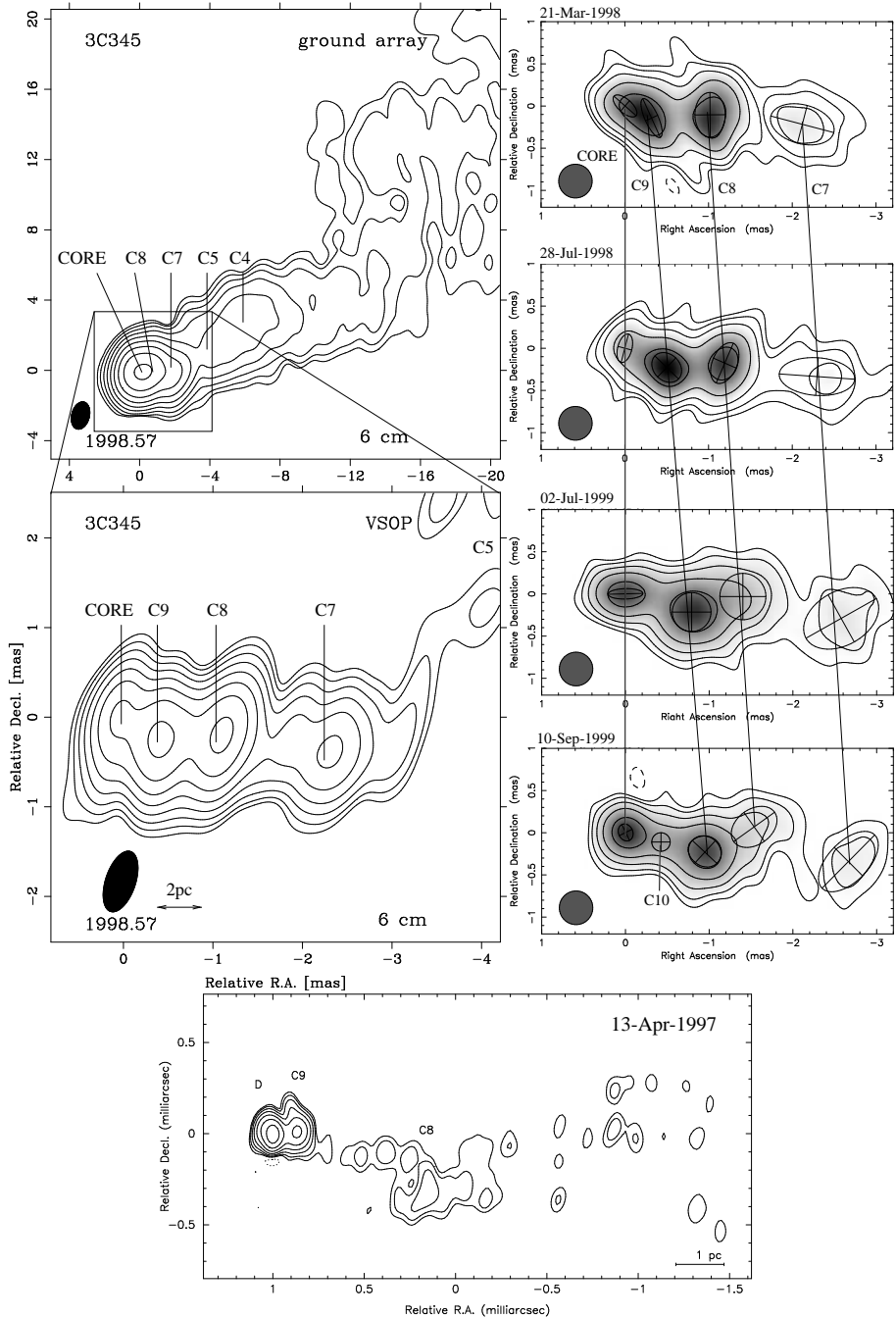,width=12cm}
\caption{Total intensity images of 3C\,345 - Top left: VSOP observations at 6\,cm resolve the nuclear region into several components. Top right: VSOP observations trace the evolution of the inner jet components at a resolution of 350$\mu$as. The ellipses are the modelfit components. Bottom: 3\,mm image of 3C\,345 with a resolution of 70$\mu$as along jet direction.}
\end{figure}

\end{document}